\documentclass[10pt,letterpaper]{article}

\usepackage[T1]{fontenc}
\usepackage[utf8]{inputenc}
\usepackage{lmodern}
\usepackage[margin=1in]{geometry}
\usepackage{amsmath,amssymb,amsfonts}
\usepackage{algorithm}
\usepackage{algorithmic}
\usepackage{graphicx}
\usepackage{xcolor}
\usepackage{booktabs}
\usepackage{multirow}
\usepackage{url}
\usepackage{hyperref}
\hypersetup{
    bookmarksopen=true,
    bookmarksnumbered=true,
    colorlinks=true,
    linkcolor=black,
    citecolor=black,
    urlcolor=blue,
    pdfstartview=FitH,
    pdfpagelayout=OneColumn
}

\title{Category-Aware Semantic Caching for Heterogeneous LLM Workloads}

\author{
\begin{tabular}{cc}
Chen Wang & Xunzhuo Liu \\
IBM Research & Tencent \\
Yorktown Heights, NY & \\
\texttt{Chen.Wang1@ibm.com} & \texttt{bitliu@tencent.com} \\[10pt]
Yue Zhu & Alaa Youssef \\
IBM Research & IBM Research \\
Yorktown Heights, NY & Yorktown Heights, NY \\
\texttt{yue.zhu@ibm.com} & \texttt{asyousse@us.ibm.com} \\[10pt]
Priya Nagpurkar & Huamin Chen \\
IBM Research & Red Hat \\
Yorktown Heights, NY & Boston, MA \\
\texttt{pnagpurkar@us.ibm.com} & \texttt{hchen@redhat.com}
\end{tabular}
}

\date{}

\begin{document}

\maketitle

\begin{abstract}
LLM serving systems process heterogeneous query workloads where different categories exhibit different characteristics. Code queries cluster densely in embedding space while conversational queries distribute sparsely. Content staleness varies from minutes (stock data) to months (code patterns). Query repetition patterns range from power-law (code) to uniform (conversation), producing long tail cache hit rate distributions: high-repetition categories achieve 40--60\% hit rates while low-repetition or volatile categories achieve 5--15\% hit rates. Vector databases must exclude the long tail because remote search costs (30ms) require 15--20\% hit rates to break even, leaving 20--30\% of production traffic uncached. Uniform cache policies compound this problem: fixed thresholds cause false positives in dense spaces and miss valid paraphrases in sparse spaces; fixed TTLs waste memory or serve stale data. This paper presents category-aware semantic caching where similarity thresholds, TTLs, and quotas vary by query category. We present a hybrid architecture separating in-memory HNSW search from external document storage, reducing miss cost from 30ms to 2ms. This reduction makes low-hit-rate categories economically viable (break-even at 3--5\% versus 15--20\%), enabling cache coverage across the entire workload distribution. Adaptive load-based policies extend this framework to respond to downstream model load, dynamically adjusting thresholds and TTLs to reduce traffic to overloaded models by 9--17\% in theoretical projections.
\end{abstract}


\section{Introduction}

LLM serving systems repeatedly execute inference on similar queries. Semantic caching returns cached responses when new queries match previous queries by semantic similarity~\cite{meancache,kvshare}. Unlike exact-match caches, semantic caches use embedding vectors and approximate nearest neighbor search~\cite{hnsw}.

Production workloads contain queries with different properties. Code generation queries use constrained vocabulary (keywords, APIs) producing dense embedding clusters. Conversational queries use varied phrasings producing sparse embeddings. Content staleness ranges from seconds (stock prices) to months (code syntax). Model costs vary by task tier---reasoning models for complex code versus smaller models for simple chat. Query repetition patterns also vary: code and documentation queries follow power-law distributions where top queries repeat frequently (40--60\% hit rates), while conversational, specialized, and volatile queries distribute uniformly with minimal repetition (5--15\% hit rates).

These differences create long tail cache hit rate distributions. In production systems, 2--3 high-volume categories (code, common documentation) achieve 40--60\% hit rates and account for 60--70\% of traffic. The remaining 30--40\% of traffic distributes across 5--10 long tail categories (specialized queries, real-time data, niche domains) achieving 5--15\% hit rates. Vector databases cannot economically serve the long tail: at 30ms remote search cost, categories need $\geq$15--20\% hit rates to break even, forcing exclusion of tail categories and leaving substantial traffic uncached.

Current semantic caches apply uniform policies: one similarity threshold, one time-to-live (TTL), equal resource allocation. This creates mismatches between policy and workload. Consider a threshold of 0.80: in dense code embeddings, this matches semantically different queries (\texttt{sort\_ascending} versus \texttt{sort\_descending}); in sparse conversational embeddings, this misses valid paraphrases. A fixed TTL of one hour either wastes space (code patterns stable for months) or serves stale data (stock prices changing per second).

This paper presents category-aware semantic caching where policies differ by query category. We identify four category properties determining appropriate policies:

\begin{enumerate}
\item Embedding space density: Vocabulary constraints affect cluster tightness
\item Query repetition patterns: Power-law versus uniform distributions
\item Content staleness rates: Update frequencies from seconds to months
\item Computational costs: Model tier pricing differences
\end{enumerate}

Vector databases cannot efficiently support per-category policies due to three architectural constraints. First, they apply similarity thresholds after completing remote search operations (typically 30ms). Per-category thresholds would require multiple searches. Second, they configure parameters at collection level rather than per-query. Third, they enforce policies server-side where category context is unavailable.

We present a hybrid architecture separating HNSW search (local, in-memory) from document storage (external database). This separation changes where policies execute: thresholds apply during graph traversal, TTL checks precede external access, and compliance validation occurs before insertion. The architecture also addresses low-hit-rate categories that vector databases must exclude. Remote search costs apply whether queries hit or miss; categories with $<$15\% hit rates become uneconomical. Local search returns immediately on miss (2ms versus 30ms remote), making lower-hit-rate categories viable.

\textbf{Contributions:}
\begin{enumerate}
\item Characterization of long tail cache hit rate distributions in heterogeneous LLM workloads and analysis of why vector databases must exclude tail categories (economic break-even at 15--20\% hit rates)
\item Category-aware caching framework where thresholds, TTLs, and quotas vary by query category based on four properties: embedding space density, repetition patterns, staleness rates, and computational costs
\item Hybrid architecture design separating in-memory HNSW search from external storage, reducing miss cost from 30ms to 2ms and lowering economic break-even to 3--5\% hit rates
\item Demonstration that hybrid architecture enables cache coverage across entire workload distribution, including long tail categories that vector databases exclude
\item Adaptive load-based policy adjustment where cache parameters respond to downstream model load, with theoretical analysis projecting 9--17\% traffic reduction to overloaded models through dynamic threshold relaxation and TTL extension
\end{enumerate}

\section{Background and Related Work}

\subsection{Semantic Caching}

Semantic caching matches queries by similarity rather than exact strings. Given query embedding $\mathbf{e}_q \in \mathbb{R}^d$, the system searches cached embeddings $\mathbf{e}_c$ where cosine similarity $\text{sim}(\mathbf{e}_q, \mathbf{e}_c) \geq \tau$ exceeds threshold $\tau$. On match, the cached response returns; otherwise the system queries the LLM and caches the result.

Recent work applies semantic caching to LLM serving. MeanCache~\cite{meancache} identifies semantically similar queries using vector databases. KVShare~\cite{kvshare} proposes semantic-aware key-value cache sharing at the LLM layer. Earlier work addressed web and mobile databases~\cite{semantic-web-queries,mobile-semantic-cache,adaptive-semantic-cache}. These systems use uniform policies across all query types.

\subsection{Vector Search}

HNSW (Hierarchical Navigable Small World)~\cite{hnsw} provides approximate nearest neighbor search with logarithmic complexity. The algorithm constructs a multi-layer graph where nodes represent vectors and edges connect similar vectors. Search starts at the top layer and traverses down, greedily moving to closer neighbors at each step.

Vector databases (Milvus~\cite{milvus}, Qdrant~\cite{qdrant}, Weaviate~\cite{weaviate}) provide distributed vector search with persistence~\cite{vdbms-survey,vdbms-bugs}. They handle billions of vectors but incur network latency. FAISS~\cite{faiss} and ScaNN~\cite{scann} optimize single-node search with GPU acceleration and quantization.

\subsection{Storage Architectures}

Systems separate hot and cold data across storage tiers. LSM-tree systems like RocksDB~\cite{rocksdb} manage write-intensive workloads with tiered compaction. Learned indexes~\cite{learned-index} replace traditional indexes with learned models predicting data locations.

\section{Category Properties and Policy Implications}

Different query categories require different cache policies. We identify four properties determining appropriate parameters.

\subsection{Embedding Space Density}

Query vocabulary constrains embedding distribution. Code queries use limited vocabulary (keywords, API names, syntax elements) producing dense clusters. The 10th nearest neighbor in code embedding spaces has distance $\approx 0.12$. Conversational queries use varied phrasings producing sparse distributions where 10th-NN distance $\approx 0.38$.

Dense spaces require tight thresholds to avoid false positives. At threshold 0.80, dense code embeddings produce 15\% false matches between semantically different queries. Threshold 0.90 reduces false matches to 3\%. Sparse spaces allow loose thresholds. At 0.80, sparse conversational embeddings miss valid paraphrases; loosening to 0.75 captures semantic equivalents without increasing false positives in the sparse space.

\subsection{Query Repetition Patterns}

Code and documentation queries follow power-law distributions (Zipfian parameter $\alpha \approx 1.2$) where common questions appear frequently~\cite{zipf-web-caching,query-workload-analysis}. Top 10\% of queries account for 45\% of traffic. Conversational queries distribute uniformly with minimal repetition.

Higher repetition justifies longer TTLs and larger quotas. Lower repetition reduces expected hit rates, warranting smaller allocations. Looser thresholds partially compensate for low repetition by capturing semantic variations.

\subsection{Content Staleness}

Content update frequency determines TTL. Code patterns and API documentation change slowly (0.01\%/day). TTLs of days or weeks avoid serving stale data while maintaining hit rates. Technical documentation updates moderately (2\%/day), suggesting TTLs of hours to a day. Stock prices, news, and weather update rapidly (80\%/hour), requiring TTLs of minutes despite reduced hit rates.

\subsection{Computational Costs}

Model costs vary by task complexity. Reasoning models (o1, GPT-4o) for code generation and complex reasoning cost more per call than smaller models (Claude 3.5 Haiku, Gemini 2.0 Flash) for simple chat. Cache hits on expensive models produce larger savings. Quota allocation can weight by economic value rather than traffic volume: allocate 40\% quota to 30\% traffic when that traffic uses expensive models.

\subsection{Long Tail Hit Rate Distribution}

The preceding properties combine to produce long tail cache hit rate distributions. Categories with power-law repetition, stable content, and dense embeddings achieve high hit rates (40--60\%). Categories with uniform repetition, volatile content, or sparse embeddings achieve low hit rates (5--15\%).

Table~\ref{tab:longtail} shows a representative production workload serving 100K queries/hour. Two head categories (code generation, API documentation) account for 60\% of traffic and achieve 45--55\% hit rates. Five tail categories account for 40\% of traffic but achieve 6--12\% hit rates due to lower repetition (conversational chat), high volatility (financial data), or specialized vocabulary (legal, medical).

\begin{table}[t]
\centering
\caption{Long tail cache hit rate distribution in production LLM serving system. Head categories achieve high hit rates; tail categories achieve low hit rates but still represent substantial traffic.}
\label{tab:longtail}
\begin{tabular}{@{}lrrrr@{}}
\toprule
Category & Traffic & Hit Rate & Vector DB & Hybrid \\
         & \%      & \%       & Viable?   & Viable? \\
\midrule
\multicolumn{5}{l}{\textit{Head (60\% traffic):}} \\
\quad Code generation & 35\% & 55\% & Yes & Yes \\
\quad API documentation & 25\% & 45\% & Yes & Yes \\
\midrule
\multicolumn{5}{l}{\textit{Tail (40\% traffic):}} \\
\quad Conversational chat & 15\% & 12\% & No & Yes \\
\quad Financial data & 10\% & 8\% & No & Yes \\
\quad Legal queries & 8\% & 10\% & No & Yes \\
\quad Medical queries & 4\% & 6\% & No & Yes \\
\quad Specialized domains & 3\% & 7\% & No & Yes \\
\bottomrule
\end{tabular}
\end{table}

Vector databases cannot economically cache tail categories. At 30ms remote search cost, break-even analysis (detailed in Section~\ref{sec:economics}) shows categories need $\geq$15--20\% hit rates. This excludes 40\% of traffic from caching, leaving substantial workload unserved. Tail categories cannot be simply ignored: they represent significant traffic volume and often serve important business use cases (financial data, healthcare).

\section{Vector Database Limitations}

Vector databases face three constraints for category-aware policies.

\subsection{Post-Search Threshold Application}

Vector databases execute similarity search remotely within the database server. Thresholds apply after search completes. Per-category thresholds require either multiple queries (30ms $\times$ N categories) or separate collections per category. Metadata filters run after search, offering no latency benefit.

\subsection{Collection-Level Configuration}

Parameters apply to entire collections rather than individual queries. Separate collections per category allow different parameters but duplicate data and prevent shared resource optimization. Multiple collections add operational complexity and infrastructure cost. Single collections with metadata filtering avoid duplication but cannot vary thresholds, TTLs, or quotas per category.

\subsection{Server-Side Policy Enforcement}

Policy enforcement occurs server-side where category context is unavailable. TTL checks require fetching documents first, wasting network calls on expired entries. Eviction operates on LRU/LFU without economic value information. Compliance validation occurs after storage, creating temporary data presence for rejected categories.

\subsection{Low-Hit-Rate Category Economics}
\label{sec:economics}

Remote search costs apply whether queries hit or miss. For vector databases, every query incurs:
\begin{itemize}
\item Network latency: 10--15ms (datacenter) or 20--30ms (cloud)
\item Vector search: 10--15ms (server-side HNSW traversal)
\item Document fetch on hit: 5--8ms (database lookup)
\end{itemize}

Total cost: approximately 30ms per query (hit or miss) plus 5ms document fetch on hit. Without caching, LLM inference takes $T_{llm}$ (e.g., 500ms for GPT-4o, 200ms for Claude 3.5 Haiku).

Break-even analysis: Let $h$ be hit rate. Expected latency with vector database caching:
\begin{equation}
L_{vdb} = 30 + h \times 5 + (1-h) \times T_{llm}
\end{equation}

Expected latency without caching: $L_{none} = T_{llm}$. Caching provides net benefit when $L_{vdb} < L_{none}$:
\begin{equation}
30 + h \times 5 + (1-h) \times T_{llm} < T_{llm}
\end{equation}

Simplifying:
\begin{equation}
30 + 5h < h \times T_{llm} \quad \Rightarrow \quad h > \frac{30}{T_{llm} - 5}
\end{equation}

For $T_{llm} = 200$ms (fast model): $h > 30/195 \approx 0.154$ (15.4\%)

For $T_{llm} = 500$ms (slow model): $h > 30/495 \approx 0.061$ (6.1\%)

Vector databases therefore require 6--15\% hit rates depending on model speed. Fast models (Claude 3.5 Haiku, Gemini 2.0 Flash) have higher break-even thresholds (15--20\%). Categories below these thresholds become uneconomical: the remote search overhead exceeds caching benefits. Production workloads contain head categories (60--70\% traffic) exceeding break-even and tail categories (30--40\% traffic) below break-even. Vector databases exclude the tail, leaving substantial traffic uncached.

\section{Hybrid Architecture}

\begin{figure}[t]
\centering
\includegraphics[width=\linewidth,keepaspectratio]{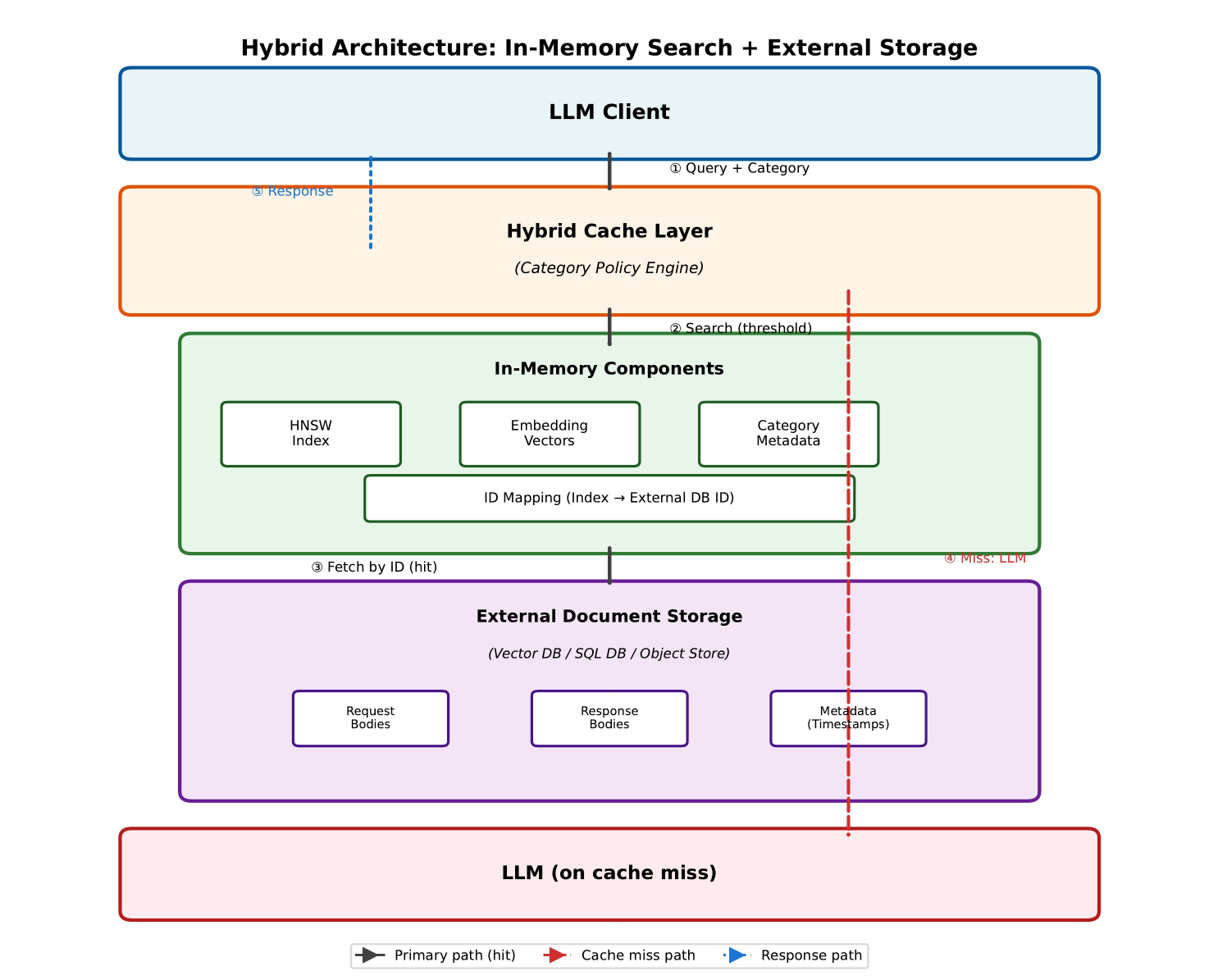}
\caption{Hybrid architecture separating in-memory HNSW search from external document storage. Query processing: (1) client submits query with category, (2) local HNSW search applies category threshold during traversal, (3) on match, fetch document by ID from external storage, (4) on miss, query LLM directly, (5) return response.}
\label{fig:architecture}
\end{figure}

The hybrid architecture separates HNSW search from document storage. Figure~\ref{fig:architecture} illustrates the design. Search executes locally in memory; documents reside in external storage (vector database, SQL database, or object store).

\subsection{Components}

The in-memory HNSW index stores graph structure, embedding vectors (1.5KB per entry for 384 dimensions), and category metadata (threshold, TTL, priority). Memory footprint: approximately 2KB per entry versus tens of kilobytes with full documents. The reduction allows larger indexes within fixed RAM capacity.

The external document store holds request bodies, response bodies, and timestamps. Access occurs by primary key lookup (5ms) rather than vector search (30ms). Storage options include vector databases used as document stores, SQL databases with ID indexing, or object stores (S3, blob storage).

The ID mapping layer connects in-memory index positions to external storage identifiers. Implementation options include hash maps or persistent key-value stores.

The category policy engine manages per-category configurations. During HNSW traversal, it applies category-specific thresholds. Before external storage access, it validates TTLs. During eviction, it considers both recency and category priority.

\subsection{Operation}

Algorithm~\ref{alg:lookup} describes cache lookup. The process: (1) retrieve category configuration, (2) check if caching is allowed for this category (compliance), (3) generate query embedding, (4) search HNSW using category threshold, (5) on miss, return NULL immediately without external access, (6) on match, validate TTL before external fetch, (7) fetch document by ID.

\begin{algorithm}[t]
\caption{Category-Aware Cache Lookup}
\label{alg:lookup}
\begin{algorithmic}[1]
\STATE \textbf{Input:} query $q$, category $c$
\STATE \textbf{Output:} cached response or NULL
\STATE
\STATE $config \gets$ getCategoryConfig($c$)
\IF{$\neg config$.allowCaching}
    \RETURN NULL
\ENDIF
\STATE
\STATE $\mathbf{e}_q \gets$ embed($q$)
\STATE $\tau \gets config$.threshold
\STATE $candidates \gets$ HNSWSearch($\mathbf{e}_q$, $\tau$)
\IF{$|candidates| = 0$}
    \RETURN NULL \COMMENT{Miss: no external access}
\ENDIF
\STATE
\STATE $idx \gets candidates[0]$
\STATE $metadata \gets$ getMetadata($idx$)
\IF{age($metadata$) $> config$.TTL}
    \STATE evict($idx$)
    \RETURN NULL
\ENDIF
\STATE
\STATE $docID \gets$ getExternalID($idx$)
\STATE $doc \gets$ fetchByID($docID$)
\RETURN $doc$.response
\end{algorithmic}
\end{algorithm}

Line 11 shows the miss path: when HNSW finds no match above threshold, return immediately. This avoids 30ms remote search that vector databases incur on misses. With 80\% miss rate, hybrid average latency is $0.2 \times 7\text{ms} + 0.8 \times 2\text{ms} = 3.0\text{ms}$ versus vector database $0.2 \times 35\text{ms} + 0.8 \times 30\text{ms} = 31\text{ms}$.

Line 14 shows TTL validation before external access. Expired entries evict immediately without wasting network calls.

\subsection{HNSW Threshold Application}

Standard k-nearest-neighbor search finds k closest vectors. Category-aware search modifies traversal: return first match exceeding category threshold rather than finding k globally nearest neighbors. This early-stopping optimization reduces latency and makes threshold application essentially free---traversal terminates on first sufficient match regardless of threshold value.

\subsection{Policy Enforcement Points}

Compliance: Categories with regulatory constraints (HIPAA, GDPR) disable caching entirely (line 5--6). Queries never enter cache, creating no temporary data presence.

TTL: Validation occurs before external access (line 14--17), avoiding network overhead on expired entries.

Quotas: Eviction considers category priority and recency. When cache is full, eviction score = priority $\times$ $1/$age $\times$ hitRate. This weights by economic value rather than pure LRU.

\subsection{Low-Hit-Rate Categories}

Shared HNSW index adds minimal marginal cost per query. Miss cost stays at 2ms whether hit rate is 5\% or 50\%. This changes break-even economics compared to vector databases.

Hybrid break-even analysis: Expected latency with hybrid caching:
\begin{equation}
L_{hybrid} = 2 + h \times 5 + (1-h) \times T_{llm}
\end{equation}

Caching provides net benefit when $L_{hybrid} < L_{none} = T_{llm}$:
\begin{equation}
2 + 5h < h \times T_{llm} \quad \Rightarrow \quad h > \frac{2}{T_{llm} - 5}
\end{equation}

For $T_{llm} = 200$ms (fast model): $h > 2/195 \approx 0.010$ (1.0\%)

For $T_{llm} = 500$ms (slow model): $h > 2/495 \approx 0.004$ (0.4\%)

Hybrid reduces break-even threshold by 15$\times$ for fast models (1\% versus 15\%) and 10$\times$ for slow models (0.4\% versus 6\%). Categories with 5--15\% hit rates that vector databases must exclude become viable in hybrid architecture.

Referring to Table~\ref{tab:longtail}: conversational chat (12\% hit rate), financial data (8\%), legal queries (10\%), medical queries (6\%), and specialized domains (7\%) all exceed hybrid break-even but fall below vector database break-even. Hybrid architecture enables cache coverage across the entire 40\% tail traffic that vector databases leave uncached.

\section{Use Cases}

\subsection{Code Generation Service (Head Category)}

Developer tools serve code generation queries. Dense embeddings require threshold 0.90 to avoid matching \texttt{sortAscending} with \texttt{sortDescending}. Power-law query distribution (top 10\% = 45\% of traffic) produces high hit rates (50--60\%). Code patterns remain stable, supporting 7-day TTL. Expensive model tier (o1, GPT-4o) justifies 40\% quota despite 30\% traffic volume. The high hit rate makes this viable for both vector databases and hybrid architecture.

\subsection{Customer Support Chatbot (Tail Category)}

Support systems serve conversational queries. Sparse embeddings allow threshold 0.75 without false positives. Uniform query distribution produces lower hit rates (10--15\%). Smaller model tier (Claude 3.5 Haiku) and lower repetition justify 15\% quota for 20\% traffic. Hit rates near or below vector database break-even (15\%) make this viable only for hybrid architecture. Vector databases would exclude this category despite representing 20\% of traffic.

\subsection{Financial Data Service (Tail Category)}

Trading platforms serve real-time queries. Content changes per second. TTL of 5 minutes prevents stale stock prices despite reduced hit rates (8--12\%). Serving stale data violates SLAs. Volatile content produces hit rates below vector database break-even. Hybrid architecture enables caching with aggressive TTL, capturing repetition within short time windows while maintaining data freshness.

\subsection{Healthcare System (Compliance-Restricted)}

Medical records contain patient queries subject to HIPAA. Configuration: \texttt{allowCaching=false}. Queries never enter cache, maintaining compliance. Other category queries use normal caching.

\section{Discussion}

\subsection{When to Use Hybrid Architecture}

Hybrid architecture benefits systems where:
\begin{itemize}
\item Workload exhibits long tail hit rate distribution
\item Cache size exceeds tens of thousands of entries
\item Documents vary in size (large responses benefit more from separation)
\item Workload contains multiple query categories with different policy requirements
\item Compliance requires pre-storage policy enforcement
\end{itemize}

If 30--40\% of traffic comes from categories with 5--15\% hit rates, vector databases will exclude this traffic while hybrid architecture makes it economically viable. Pure in-memory caches remain simpler for small datasets ($<$10K entries). Pure vector databases handle workloads where network latency matters less than operational simplicity, or where all categories achieve high hit rates ($>$20\%).

\subsection{Category Classification}

Categories can be determined by:
\begin{itemize}
\item Explicit routing: Client specifies category in request metadata
\item Endpoint-based: Different API endpoints serve different categories
\item Prompt classification: Lightweight classifier routes queries (adds latency)
\end{itemize}

Explicit routing and endpoint-based approaches add zero classification overhead.

\subsection{Policy Tuning}

Initial policies derive from category properties: dense spaces use tight thresholds ($\geq$0.88), sparse spaces use loose thresholds ($\leq$0.78); stable content uses long TTLs ($\geq$3 days), volatile content uses short TTLs ($\leq$10 minutes); expensive models receive larger quotas.

Refinement uses A/B testing: route portion of traffic to alternate threshold, measure hit rate and accuracy, adopt winning configuration. Alternatively, automated adaptation adjusts policies based on observed staleness rates and hit rates.

\subsection{Memory Scaling}

HNSW search complexity: $O(\log n)$ where $n$ is index size. Practical latencies: 2--3ms for 1M entries, 5--8ms for 10M entries. Beyond 10M entries, consider sharding by category or vector space region.

Memory overhead beyond embeddings: ID mappings (16 bytes/entry), category metadata (64 bytes/entry), statistics (32 bytes/entry). Total overhead $\approx$112 bytes per entry ($\sim$5\% of 2KB baseline).

\subsection{Adaptive Load-Based Policies}

Routing systems direct queries to multiple models with varying loads. When a model experiences high traffic, query latency increases and queues grow. Cache policies can adapt to model load: relax similarity thresholds and extend TTLs when downstream models are busy, tighten policies when models are idle.

\subsubsection{Model Load and Cache Value}

Model latency under load changes cache economics. Let $T_{base}$ be normal model latency and $T_{load} = \alpha \cdot T_{base}$ be latency under load where $\alpha > 1$ represents load multiplier. Common production scenarios: $\alpha = 2$--4 during traffic spikes, $\alpha = 1.5$--2 during sustained high utilization.

Cache hit value increases with model latency. Without caching, user experiences $T_{load}$. With caching, user experiences cache latency $T_{cache}$ on hit. Savings per hit: $T_{load} - T_{cache}$. Higher $\alpha$ increases per-hit value, justifying more aggressive caching.

Break-even analysis under load: Hybrid architecture provides benefit when expected latency with caching is less than without:
\begin{equation}
2 + h \times 5 + (1-h) \times T_{load} < T_{load}
\end{equation}

This simplifies to $h > 2/(T_{load} - 5)$. For loaded model with $T_{load} = 1000$ms (slow model under 2$\times$ load): $h > 2/995 \approx 0.002$ (0.2\%). Even 1\% hit rate provides net benefit under high load.

\subsubsection{Threshold Relaxation}

Decreasing similarity threshold increases hit rate by matching more distant embeddings. Let base threshold $\tau_0$ produce hit rate $h_0$. Relaxed threshold $\tau_1 = \tau_0 - \delta$ produces hit rate $h_1 = h_0 + \Delta h$ where $\Delta h$ depends on embedding space density and query distribution.

Traffic reduction to model: Queries reaching model decrease from $(1-h_0)$ to $(1-h_1) = (1-h_0-\Delta h)$. Load reduction factor: $\Delta h / (1-h_0)$. Example: category with 40\% base hit rate ($h_0 = 0.40$) serves 60\% traffic to model. Increasing hit rate to 50\% ($\Delta h = 0.10$) reduces model traffic by $0.10/0.60 \approx 16.7\%$.

Threshold adjustment has limits. False positive rate increases with relaxation. Dense embedding spaces (code, APIs) tolerate smaller adjustments ($\delta \leq 0.05$) before matching semantically distinct queries. Sparse spaces (conversation) tolerate larger adjustments ($\delta \leq 0.10$) without significant false positives.

\subsubsection{TTL Extension}

Extending TTL keeps entries longer, increasing hit rate for repeated queries. Let base TTL $t_0$ with staleness rate $s$ (fraction of content changing per unit time). Extended TTL $t_1 = \beta \cdot t_0$ where $\beta > 1$ increases stale response probability from $s \cdot t_0$ to $s \cdot t_1 = \beta \cdot s \cdot t_0$.

Trade-off: longer TTL improves hit rates but increases staleness. Acceptable when model load ($\alpha$) creates worse user experience than moderate staleness. Consider stock price queries with $t_0 = 5$ minutes and $s = 0.20$ (20\% change per 5 minutes). Base staleness: 20\%. Extending to $t_1 = 15$ minutes ($\beta = 3$): staleness increases to 60\%. This remains acceptable if alternative is 2000ms latency versus 100ms with potentially stale data.

Categories with low staleness rates benefit most from TTL extension. Code patterns with $s = 0.0001$ (0.01\% daily change rate) support large $\beta$ values. Extending TTL from 3 days to 9 days increases staleness from 0.3\% to 0.9\%---negligible for most applications.

\subsubsection{Dynamic Adjustment Strategy}

Policy adjustment based on observed model metrics. Let $L_p$ be model latency percentile (e.g., P95) and $Q$ be queue depth. Define load factor:
\begin{equation}
\lambda = \min\left(1, \frac{L_p}{L_{target}} \cdot w_L + \frac{Q}{Q_{target}} \cdot w_Q\right)
\end{equation}
where $L_{target}$ and $Q_{target}$ are acceptable thresholds, $w_L$ and $w_Q$ are weights ($w_L + w_Q = 1$).

Current threshold: $\tau(\lambda) = \tau_0 - \lambda \cdot \delta_{max}$ where $\delta_{max}$ is maximum relaxation. Current TTL: $t(\lambda) = t_0 \cdot (1 + \lambda \cdot (\beta_{max} - 1))$ where $\beta_{max}$ is maximum extension factor.

Example configuration: Code generation category with $\tau_0 = 0.90$, $\delta_{max} = 0.05$, $t_0 = 7$ days, $\beta_{max} = 2$. Under normal load ($\lambda = 0$): threshold 0.90, TTL 7 days. Under high load ($\lambda = 1$): threshold 0.85, TTL 14 days.

Adjustment reduces model traffic. Assume hit rate increases linearly with threshold relaxation: $\Delta h = k \cdot \delta$ where $k$ is category-specific sensitivity (typical range: $k = 0.5$--2.0 per 0.01 threshold change). For $k = 1.0$, relaxing by $\delta = 0.05$ increases hit rate by $\Delta h = 0.05$ (5 percentage points). Category with base 45\% hit rate reaches 50\%, reducing model traffic from 55\% to 50\%---a 9\% reduction in model load.

\textbf{Note:} The traffic reduction figures (9--17\%) are theoretical projections based on assumed linear relationships between threshold relaxation and hit rate improvements. Actual performance depends on workload-specific characteristics including query distribution, embedding space geometry, and category-specific sensitivity parameters. Empirical validation with production workloads is needed to confirm these projections.

\subsubsection{Multi-Model Routing}

Routing systems distribute queries across models with independent loads. Per-model policies adapt independently. Model A under load relaxes threshold while Model B under light load maintains tight threshold. This distributes cache resources toward loaded models.

Consider two models: Model A (o1) serves 30\% traffic at \$0.10/call with 500ms latency; Model B (GPT-4o-mini) serves 70\% traffic at \$0.01/call with 150ms latency. Under normal load, both use base thresholds. When Model A experiences 3$\times$ load spike (1500ms latency), its policy relaxes while Model B maintains base policy. Cache hits on Model A now save 1500ms and \$0.10 versus 150ms and \$0.01 on Model B---10$\times$ latency savings and 10$\times$ cost savings. Preferential caching of loaded, expensive models maximizes system-wide benefit.

\subsubsection{Implementation Considerations}

Adaptation requires damping to prevent oscillation. Moving average over 5--10 minute windows smooths transient spikes. Hysteresis prevents rapid threshold changes: require load factor to change by $\geq$0.1 before adjusting policy.

Safety bounds prevent excessive relaxation. Minimum threshold prevents false positive rates above acceptable levels (e.g., $\tau_{min} = 0.80$ for dense spaces). Maximum TTL caps staleness (e.g., $t_{max} = 2 \times t_0$ for moderate staleness tolerance).

Monitoring tracks false positive rates during relaxed periods. If false positive rate exceeds threshold (e.g., 5\%), reduce $\delta_{max}$. This provides feedback loop between observed accuracy and policy bounds.

\subsection{Extensions}

Document caching: Store subset of hot documents in memory alongside HNSW. Top 5\% of entries account for 80\% of hits (power law)~\cite{zipf-web-caching}. In-memory document access reduces hit latency from 7ms to 2ms.

Compression: Compress documents in external storage using zstd (60--70\% reduction, 2ms decompression) or lz4 (40--50\% reduction, 0.5ms decompression).

Multi-tier: L1 in-memory with hot documents (10K entries, $<$2ms), L2 local SSD with HNSW (100K entries, $<$10ms), L3 external database (unlimited, $<$50ms).

\section{Related Work}

\subsection{LLM Serving and Caching}

Recent work optimizes LLM inference through various caching mechanisms. MeanCache~\cite{meancache} introduces user-centric semantic caching using vector databases. KVShare~\cite{kvshare} proposes semantic-aware cache sharing at the key-value layer within LLM inference. VaryGen~\cite{varygen} generates test inputs for semantic cache evaluation. These systems apply uniform policies across query types and rely on vector database architectures. They do not characterize long tail hit rate distributions in heterogeneous workloads or address the economic constraints that force vector databases to exclude low-hit-rate categories. Our work identifies this fundamental limitation and presents an architecture that makes tail categories economically viable.

\subsection{Database Semantic Caching}

Early semantic caching addressed web and mobile databases. Chidlovskii and Borghoff~\cite{semantic-web-queries} developed semantic caching for web queries using signature files. Shi et al.~\cite{mobile-semantic-cache} proposed cache management for mobile databases. Limam and Akaichi~\cite{adaptive-semantic-cache} introduced adaptive replacement algorithms. These works establish semantic similarity principles but do not address heterogeneous workload optimization.

\subsection{Vector Databases}

Pan et al.~\cite{vdbms-survey} survey vector database systems, analyzing over 20 commercial implementations. Xie et al.~\cite{vdbms-bugs} study software defects in vector databases. Milvus~\cite{milvus}, Qdrant~\cite{qdrant}, and Weaviate~\cite{weaviate} provide production vector search systems. These systems couple indexing and storage, making per-query policy customization expensive.

\subsection{Approximate Nearest Neighbor Search}

Malkov and Yashunin~\cite{hnsw} introduced HNSW graphs achieving logarithmic search complexity. FAISS~\cite{faiss} provides GPU-optimized similarity search with quantization. ScaNN~\cite{scann} uses anisotropic vector quantization. These systems focus on search performance; our work addresses policy heterogeneity in caching applications.

\section{Conclusion}

This paper presents category-aware semantic caching for heterogeneous LLM workloads. Production systems exhibit long tail cache hit rate distributions: 2--3 head categories achieve 40--60\% hit rates and account for 60--70\% of traffic, while 5--10 tail categories achieve 5--15\% hit rates but represent 30--40\% of traffic. Different query categories require different cache policies based on embedding space density, repetition patterns, staleness rates, and computational costs.

Vector databases cannot efficiently support heterogeneous workloads for two reasons. First, they cannot efficiently implement per-category policies because they apply thresholds after remote search, configure parameters collection-wide, and enforce policies server-side. Second, they cannot economically serve long tail categories: remote search costs (30ms) require 6--15\% hit rates to break even, forcing exclusion of tail categories below this threshold.

The hybrid architecture separates in-memory HNSW search from external document storage. This separation allows threshold application during local graph traversal, TTL validation before external access, and compliance enforcement before insertion. It reduces miss cost from 30ms (remote search) to 2ms (local return), lowering break-even threshold by 10--15$\times$ to 0.4--1.0\%. Categories with 5--15\% hit rates that vector databases must exclude become viable, enabling cache coverage across the entire workload distribution.

Adaptive load-based policies extend the framework to respond to downstream model load. When models experience traffic spikes or sustained high utilization, cache policies relax thresholds and extend TTLs to increase hit rates and reduce model traffic. Theoretical analysis projects threshold relaxation of 0.05 can reduce model traffic by 9--17\% depending on base hit rate, though empirical validation is needed. Per-model adaptation directs cache resources toward loaded models, particularly valuable when expensive models experience load while cheaper models remain idle.

The approach applies to LLM serving systems processing diverse query categories where long tail distributions and per-category policy requirements make vector database architectures insufficient.

\section*{Acknowledgments}

This work was supported by vLLM community.

\bibliographystyle{plain}

\end{document}